\documentclass[amsmath,amssymb,aps, twocolumn]{aastex62}

\usepackage{amsmath}
\usepackage{amssymb}
\usepackage{verbatim}
\usepackage{enumerate}
\usepackage{graphicx}
\usepackage{bm}
\usepackage{microtype}

\newcommand{\indii}{DES J2038--4609}
\newcommand{\sref}{\S\ref}
\graphicspath{{./}{figures/}}

\received{May 8, 2020}
\submitjournal{ApJ}

\shorttitle{Grus I \&  Indus II}
\shortauthors{Cantu et al.}

\begin{document}

\title{\bf\large A Deeper Look at DES Dwarf Galaxy Candidates: Grus I and  Indus II}

\correspondingauthor{Sarah A. Cantu}
\email{scantu3@tamu.edu}

\author[0000-0001-6576-4173]{Sarah A. Cantu}
\affiliation{George P. and Cynthia Woods Mitchell Institute for Fundamental Physics and Astronomy, Texas A\&M University, College Station, TX 77843, USA}
\affiliation{Department of Physics \& Astronomy, Texas A\&M University, 4242 TAMU, College Station, TX 77843, USA}

\author{Andrew B. Pace}
\affiliation{McWilliams Center for Cosmology, Carnegie Mellon University, 5000 Forbes Ave., Pittsburgh, PA 15213, USA}
\affiliation{George P. and Cynthia Woods Mitchell Institute for Fundamental Physics and Astronomy, Texas A\&M University, College Station, TX 77843, USA}
\affiliation{Department of Physics \& Astronomy, Texas A\&M University, 4242 TAMU, College Station, TX 77843, USA}

\author{Jennifer Marshall}
\affiliation{George P. and Cynthia Woods Mitchell Institute for Fundamental Physics and Astronomy, Texas A\&M University, College Station, TX 77843, USA}
\affiliation{Department of Physics \& Astronomy, Texas A\&M University, 4242 TAMU, College Station, TX 77843, USA}
%\collaboration{(Dark Energy Survey)}

\author{Louis E. Strigari}
\affiliation{George P. and Cynthia Woods Mitchell Institute for Fundamental Physics and Astronomy, Texas A\&M University, College Station, TX 77843, USA}
\affiliation{Department of Physics \& Astronomy, Texas A\&M University, 4242 TAMU, College Station, TX 77843, USA}

\author{Denija Crnojevic}
\affiliation{Department of Chemistry and Physics, University of Tampa,
Tampa, FL, USA}

\author{Joshua D. Simon}
\affiliation{Observatories of the Carnegie Institution for Science, 813 Santa Barbara St., Pasadena, CA 91101, USA}

\author{A.~Drlica-Wagner}
\affiliation{Fermi National Accelerator Laboratory, P.O. Box 500, Batavia, IL 60510, USA}
\affiliation{Kavli Institute for Cosmological Physics, University of Chicago, Chicago, IL 60637, USA}
\affiliation{Department of Astronomy and Astrophysics, University of Chicago, Chicago, IL 60637, USA}

\author{K.~Bechtol}
\affiliation{Physics Department, 2320 Chamberlin Hall, University of Wisconsin-Madison, 1150 University Avenue Madison, WI 53706-1390}
\affiliation{LSST, 933 North Cherry Avenue, Tucson, AZ 85721, USA}

\author{Clara E. Mart\'{i}nez-V\'{a}zquez}
\affiliation{Cerro Tololo Inter-American Observatory, NSF's National Optical-Infrared Astronomy Research Laboratory, Casilla 603, La Serena, Chile}

\author{B. Santiago}
\affiliation{Instituto de F\'{i}sica, UFRGS, caixa Postal 15051, Porto Alegre, RS-91501-970, Brazil}
\affiliation{Laborat\'{o}rio Interinstitucional de e-Astronomia - LIneA, Rua Gal. Jos\'{e} Cristino 77, Rio de Janeiro, RJ - 20921-400, Brazil}

\author{A.~Amara}
\affiliation{Department of Physics, ETH Zurich, Wolfgang-Pauli-Strasse 16, CH-8093 Zurich, Switzerland} 
\author{K. M. Stringer}
\affiliation{George P. and Cynthia Woods Mitchell Institute for Fundamental Physics and Astronomy, Texas A\&M University, College Station, TX 77843, USA}
\affiliation{Department of Physics \& Astronomy, Texas A\&M University, 4242 TAMU, College Station, TX 77843, USA}
\author{H.~T.~Diehl}
\affiliation{Fermi National Accelerator Laboratory, P. O. Box 500, Batavia, IL 60510, USA}

%%%%% BUILDERS %%%%%%%%%%%%%

\author{M.~Aguena}
\affiliation{Departamento de F\'isica Matem\'atica, Instituto de F\'isica, Universidade de S\~ao Paulo, CP 66318, S\~ao Paulo, SP, 05314-970, Brazil}
\affiliation{Laborat\'orio Interinstitucional de e-Astronomia - LIneA, Rua Gal. Jos\'e Cristino 77, Rio de Janeiro, RJ - 20921-400, Brazil}
\author{S.~Allam}
\affiliation{Fermi National Accelerator Laboratory, P. O. Box 500, Batavia, IL 60510, USA}
\author{S.~Avila}
\affiliation{Instituto de Fisica Teorica UAM/CSIC, Universidad Autonoma de Madrid, 28049 Madrid, Spain}
\author{D.~Brooks}
\affiliation{Department of Physics \& Astronomy, University College London, Gower Street, London, WC1E 6BT, UK}
\author{A.~Carnero~Rosell}
\affil{Laborat\'orio Interinstitucional de e-Astronomia - LineA, Rua Gal. Jos\'e Cristino 77, Rio de Janeiro, RJ - 20921-400, Brazil}
\affil{Observat\'orio Nacional, Rua Gal. Jos\'e Cristino 77, Rio de Janeiro, RJ - 20921-400, Brazil}
\author{M.~Carrasco~Kind}
\affiliation{Department of Astronomy, University of Illinois at Urbana-Champaign, 1002 W. Green Street, Urbana, IL 61801, USA}
\affiliation{National Center for Supercomputing Applications, 1205 West Clark St., Urbana, IL 61801, USA}
\author{J.~Carretero}
\affiliation{Institut de F\'{\i}sica d'Altes Energies (IFAE), The Barcelona Institute of Science and Technology, Campus UAB, 08193 Bellaterra (Barcelona) Spain}
\author{M.~Costanzi}
\affiliation{INAF-Osservatorio Astronomico di Trieste, via G. B. Tiepolo 11, I-34143 Trieste, Italy}
\affiliation{Institute for Fundamental Physics of the Universe, Via Beirut 2, 34014 Trieste, Italy}
\author{L.~N.~da Costa}
\affiliation{Laborat\'orio Interinstitucional de e-Astronomia - LIneA, Rua Gal. Jos\'e Cristino 77, Rio de Janeiro, RJ - 20921-400, Brazil}
\affiliation{Observat\'orio Nacional, Rua Gal. Jos\'e Cristino 77, Rio de Janeiro, RJ - 20921-400, Brazil}
\author{J.~De~Vicente}
\affiliation{Centro de Investigaciones Energ\'eticas, Medioambientales y Tecnol\'ogicas (CIEMAT), Madrid, Spain}
\author{S.~Desai}
\affiliation{Department of Physics, IIT Hyderabad, Kandi, Telangana 502285, India}
\author{P.~Doel}
\affiliation{Department of Physics \& Astronomy, University College London, Gower Street, London, WC1E 6BT, UK}
\author{T.~F.~Eifler}
\affiliation{Department of Astronomy/Steward Observatory, University of Arizona, 933 North Cherry Avenue, Tucson, AZ 85721-0065, USA}
\affiliation{Jet Propulsion Laboratory, California Institute of Technology, 4800 Oak Grove Dr., Pasadena, CA 91109, USA}
\author{S.~Everett}
\affiliation{Santa Cruz Institute for Particle Physics, Santa Cruz, CA 95064, USA}
\author{J.~Frieman}
\affiliation{Fermi National Accelerator Laboratory, P. O. Box 500, Batavia, IL 60510, USA}
\affiliation{Kavli Institute for Cosmological Physics, University of Chicago, Chicago, IL 60637, USA}
\author{J.~Garc\'{i}a-Bellido}
\affiliation{Instituto de Fisica Teorica UAM/CSIC, Universidad Autonoma de Madrid, 28049 Madrid, Spain}
\author{E.~Gaztanaga}
\affiliation{Institut d'Estudis Espacials de Catalunya (IEEC), 08034 Barcelona, Spain}
\affiliation{Institute of Space Sciences (ICE, CSIC),  Campus UAB, Carrer de Can Magrans, s/n,  08193 Barcelona, Spain}
\author{D.~Gruen}
\affiliation{Department of Physics, Stanford University, 382 Via Pueblo Mall, Stanford, CA 94305, USA}
\affiliation{Kavli Institute for Particle Astrophysics \& Cosmology, P. O. Box 2450, Stanford University, Stanford, CA 94305, USA}
\affiliation{SLAC National Accelerator Laboratory, Menlo Park, CA 94025, USA}
\author{R.~A.~Gruendl}
\affiliation{Department of Astronomy, University of Illinois at Urbana-Champaign, 1002 W. Green Street, Urbana, IL 61801, USA}
\affiliation{National Center for Supercomputing Applications, 1205 West Clark St., Urbana, IL 61801, USA}
\author{J.~Gschwend}
\affiliation{Laborat\'orio Interinstitucional de e-Astronomia - LIneA, Rua Gal. Jos\'e Cristino 77, Rio de Janeiro, RJ - 20921-400, Brazil}
\affiliation{Observat\'orio Nacional, Rua Gal. Jos\'e Cristino 77, Rio de Janeiro, RJ - 20921-400, Brazil}
\author{G.~Gutierrez}
\affiliation{Fermi National Accelerator Laboratory, P. O. Box 500, Batavia, IL 60510, USA}
\author{S.~R.~Hinton}
\affiliation{School of Mathematics and Physics, University of Queensland,  Brisbane, QLD 4072, Australia}
\author{D.~L.~Hollowood}
\affiliation{Santa Cruz Institute for Particle Physics, Santa Cruz, CA 95064, USA}
\author{K.~Honscheid}
\affiliation{Center for Cosmology and Astro-Particle Physics, The Ohio State University, Columbus, OH 43210, USA}
\affiliation{Department of Physics, The Ohio State University, Columbus, OH 43210, USA}
\author{D.~J.~James}
\affiliation{Center for Astrophysics $\vert$ Harvard \& Smithsonian, 60 Garden Street, Cambridge, MA 02138, USA}
\author{K.~Kuehn}
\affiliation{Australian Astronomical Optics, Macquarie University, North Ryde, NSW 2113, Australia}
\affiliation{Lowell Observatory, 1400 Mars Hill Rd, Flagstaff, AZ 86001, USA}
\author{M.~A.~G.~Maia}
\affiliation{Laborat\'orio Interinstitucional de e-Astronomia - LIneA, Rua Gal. Jos\'e Cristino 77, Rio de Janeiro, RJ - 20921-400, Brazil}
\affiliation{Observat\'orio Nacional, Rua Gal. Jos\'e Cristino 77, Rio de Janeiro, RJ - 20921-400, Brazil}
\author{F.~Menanteau}
\affiliation{Department of Astronomy, University of Illinois at Urbana-Champaign, 1002 W. Green Street, Urbana, IL 61801, USA}
\affiliation{National Center for Supercomputing Applications, 1205 West Clark St., Urbana, IL 61801, USA}
\author{R.~Miquel}
\affiliation{Instituci\'o Catalana de Recerca i Estudis Avan\c{c}ats, E-08010 Barcelona, Spain}
\affiliation{Institut de F\'{\i}sica d'Altes Energies (IFAE), The Barcelona Institute of Science and Technology, Campus UAB, 08193 Bellaterra (Barcelona) Spain}
\author{A.~Palmese}
\affiliation{Fermi National Accelerator Laboratory, P. O. Box 500, Batavia, IL 60510, USA}
\affiliation{Kavli Institute for Cosmological Physics, University of Chicago, Chicago, IL 60637, USA}
\author{F.~Paz-Chinch\'{o}n}
\affiliation{Institute of Astronomy, University of Cambridge, Madingley Road, Cambridge CB3 0HA, UK}
\affiliation{National Center for Supercomputing Applications, 1205 West Clark St., Urbana, IL 61801, USA}
\author{A.~A.~Plazas}
\affiliation{Department of Astrophysical Sciences, Princeton University, Peyton Hall, Princeton, NJ 08544, USA}
\author{E.~Sanchez}
\affiliation{Centro de Investigaciones Energ\'eticas, Medioambientales y Tecnol\'ogicas (CIEMAT), Madrid, Spain}
\author{V.~Scarpine}
\affiliation{Fermi National Accelerator Laboratory, P. O. Box 500, Batavia, IL 60510, USA}
\author{M.~Schubnell}
\affiliation{Department of Physics, University of Michigan, Ann Arbor, MI 48109, USA}
\author{S.~Serrano}
\affiliation{Institut d'Estudis Espacials de Catalunya (IEEC), 08034 Barcelona, Spain}
\affiliation{Institute of Space Sciences (ICE, CSIC),  Campus UAB, Carrer de Can Magrans, s/n,  08193 Barcelona, Spain}
\author{I.~Sevilla-Noarbe}
\affiliation{Centro de Investigaciones Energ\'eticas, Medioambientales y Tecnol\'ogicas (CIEMAT), Madrid, Spain}
\author{M.~Smith}
\affiliation{School of Physics and Astronomy, University of Southampton,  Southampton, SO17 1BJ, UK}
\author{M.~Soares-Santos}
\affiliation{Brandeis University, Physics Department, 415 South Street, Waltham MA 02453}
\author{E.~Suchyta}
\affiliation{Computer Science and Mathematics Division, Oak Ridge National Laboratory, Oak Ridge, TN 37831}
\author{M.~E.~C.~Swanson}
\affiliation{National Center for Supercomputing Applications, 1205 West Clark St., Urbana, IL 61801, USA}
\author{G.~Tarle}
\affiliation{Department of Physics, University of Michigan, Ann Arbor, MI 48109, USA}
\author{A.~R.~Walker}
\affiliation{Cerro Tololo Inter-American Observatory, National Optical Astronomy Observatory, Casilla 603, La Serena, Chile}
\author{R.D.~Wilkinson}
\affiliation{Department of Physics and Astronomy, Pevensey Building, University of Sussex, Brighton, BN1 9QH, UK}
\collaboration{(DES Collaboration)}

\begin{abstract}

We present deep $g$- and $r$-band Magellan/Megacam photometry of two dwarf galaxy candidates discovered in the Dark Energy Survey (DES), Grus I and Indus II (\indii).  For the case of Grus I, we resolved the main sequence turn-off (MSTO) and $\sim 2$ mags below it. The MSTO can be seen at $g_0\sim 24$ with a photometric uncertainty of $0.03$ mag. We show Grus I to be consistent with an old, metal-poor ($\sim 13.3$ Gyr, [Fe/H]$\sim-1.9$) dwarf galaxy.  We derive updated distance and structural parameters for Grus I using this deep, uniform, wide-field data set.  We find an azimuthally averaged half-light radius more than two times larger ($\sim 151^{+21}_{-31}$ pc; $\sim 4\farcm16^{+0.54}_{-0.74}$) and an absolute $V$-band magnitude $\sim-4.1$ that is $\sim 1$ magnitude brighter than previous studies. We obtain updated distance, ellipticity, and centroid parameters which are in agreement with other studies within uncertainties. Although our photometry of Indus II is $\sim 2-3$ magnitudes deeper than the DES Y1 Public release, we find no coherent stellar population at its reported location. The original detection was located in an incomplete region of sky in the DES Y2Q1 data set and was flagged due to potential blue horizontal branch member stars. The best fit isochrone parameters are physically inconsistent with both dwarf galaxies and globular clusters. We conclude that Indus II is likely a false-positive, flagged due to a chance alignment of stars along the line of sight.

\end{abstract}

\keywords{galaxies: dwarf --- galaxies: fundamental parameters --- methods: data analysis --- methods: statistical --- techniques: photometric}

\reportnum{DES-2018-0422}
\reportnum{FERMILAB-PUB-20-188-V}

\section{Introduction}\label{sec:intro}

With the advent of high-precision, large-area surveys such as SDSS \citep{York2000TheSummary}, the Dark Energy Survey \citep[DES;][]{Abbott2005TheSurvey}, Pan-STARRS \citep{Chambers2016TheSurveys}, the Hyper Suprime-Cam Subaru Strategic Program \citep[HSC SSP;][]{Homma2016}, MagLites \citep{2016ApJ...833L...5D, maglitescar2_3}, and DELVE \citep{mau2019}, the number of known faint satellite systems that orbit the Milky Way (MW) has dramatically increased \citep{TheDESCollaboration2015, Koposov2015, Bechtol2015}. The ambiguity of what constitutes a galaxy increases as more systems are discovered that lie between the traditional loci of globular clusters and galaxies.  Additionally, these low-luminosity systems challenge spectroscopic studies due to their low number of bright member stars \citep{Willman2012GalaxyDefined}.

Many of these satellites discovered in the past decade are categorized as ultra-faint dwarf (UFD) galaxies~\citep{2019arXiv190105465S}. With $M_V \ga -8$ mag \citep[$M_* \la 10^5 M_\sun$;][]{Martin2008, Bullock2017}, UFDs overlap with bright globular clusters (GCs) in the size-luminosity plane. Though they overlap in this parameter space, UFDs and GCs likely have different formation mechanisms~\citep{2018RSPSA.47470616F}. From their internal stellar kinematics, GCs are consistent with having little or no dark matter, and may be remnants of nucleated dwarf galaxies or may follow a completely separate evolutionary path \citep{nuclateddwarfs}.

In contrast, the stellar kinematics of UFDs exhibit high $M/L_V$ ratios \cite[i.e., $M/L_V\sim 10^3$;][]{Simon2007} and represent the faintest end of the galaxy luminosity function. Dynamical mass measurements are one of the primary distinguishing characteristics between UFDs and GCs. 
In comparison to low-luminosity GCs, UFDs have larger sizes ($r_h\ga 30$ pc), larger velocity dispersions ($\sigma\gtrsim 3$ km s$^{-1}$), and significant metallicity spreads ($\sigma_{[Fe/H]}\ga 0.3$ dex), as shown in \cite{Simon2007} and \cite{2007MNRAS.380..281M}.

As the most dark matter dominated objects visible in the Universe, UFDs provide crucial, empirical information about the nature of dark matter and hierarchical structure at the smallest-scales~\citep{2012AnP...524..507F, Bullock2017}. In $\Lambda$-cold dark matter ($\Lambda$CDM) cosmology, structure forms hierarchically, with the UFDs corresponding to the galaxies in the smallest of dark matter halos \citep{2015MNRAS.453.1305W,Sawala2015BentRelation, Wetzel2016}. Discerning the exact nature of MW satellites is therefore our paramount observational method to better constrain and compare cosmological models to low-luminosity systems. Firmly establishing the newly-discovered satellites as UFDs, and measuring their mass-to-light ratios, requires spectroscopic studies of a significant sample of their stars \citep[e.g.,][]{Li2018}.

However, due to the faintness of these systems, spectroscopy is only possible for a small sample of their stars, making a robust determination of their mass-to-light ratios difficult to obtain.
In addition to spectroscopic studies, information on the structural parameters and stellar populations of UFDs may be obtained through deep photometric studies. 
For faint overdensities of stars like UFDs, this requires targeted imaging and precise photometry, in order to distinguish members of the systems from background stars and galaxies~\citep[e.g.,][]{Martin2008,Muoz2012MeasuringGalaxies,Brown2014,Conn2017OnV,2018ApJ...857...70C,2018arXiv180902259J, mutlu2018}.

In this work we seek to clarify the nature of two objects detected in the DES footprint, Grus I and Indus II, with deep Magellan/Megacam imaging. 

Grus I was discovered by \citet{Koposov2015}, however its status as a GC or UFD has not yet been totally disentangled due to its faintness ($M_V = -3.4$) and the lack of deep, wide field photometry. Follow-up studies based on the deep but small Gemini/GMOS-S field of view (FOV) photometry \citep{2018arXiv180902259J} were not able to determine the properties of Grus I because of its extension \citep[$r_h = 1\farcm77$,][]{Koposov2015}. 

\citet{martinez19} obtained a precise distance to Grus I of $D_{\odot} = 127 \pm 6$ kpc ($\mu_0 = 20.51 \pm 0.10 $ mag) from the detection of two RR Lyrae members.  They find that this distance would imply a change of 5\% in its previously calculated physical size, consistent with the estimate of \citet{Koposov2015}. Given the large uncertainties in the previous determinations of physical size, deep and extended imaging in Grus I is needed to firmly confirm this.

Complementary spectroscopic studies made of this system \citep{Walker2016, 2019ApJ...870...83J} were not able to decipher the nature of this object either, since the velocity dispersion could not be resolved because of the scarce sample of members detected. 

Our second target, \indii, was identified in \citet{TheDESCollaboration2015} as a low-confidence UFD candidate and will be referred to as Indus II throughout the paper for convenience. The initial data for Indus II were located in a survey region with atypical non-uniformity as they were taken part-way through the survey observations. The primary evidence for candidacy stems from a clump of apparent blue horizontal branch (BHB) stars at $g\sim 22$. While Indus II has been targeted in some dark matter indirect detection analyses \citep{Albert2017}, there are no other studies confirming the nature of the object. Given the uncertainty associated with this system, we chose to confirm whether this target was a  gravitationally bound system due to Magellan/Megacam's FOV potentially covering $3\times r_h$ \citep{TheDESCollaboration2015}. 

We follow similar methods to other studies that have confirmed the status of many MW satellites as dwarf galaxies \citep[see, e.g.,][]{Sand2012,2016ApJ...824L..14C,2016ApJ...833...16K,2016MNRAS.458..603L,2017AJ....154..267C,Conn2017OnV,2018ApJ...857...70C,Luque2018,2018AAS...23141203M}. Our data complements other studies by utilizing a larger FOV (necessary for the potentially larger extents), while still resolving magnitudes $\sim 3$ magnitudes deeper than the discovery papers.

This paper is structured as follows. In~\sref{sec:obs} we describe the Magellan/Megacam and DES observations, photometry, and catalog selection.  We present the likelihood method used to infer structural parameters in \sref{sec:Anal}. In \sref{sec:Res} we report the results from the statistical analysis, and \sref{sec:disc} compares our results to previous results and concludes this work.

\section{Observations and Data Reduction} \label{sec:obs}

\begin{figure*}[!t]
\centering
\begin{tabular}{c}
  \includegraphics[width=1\linewidth]{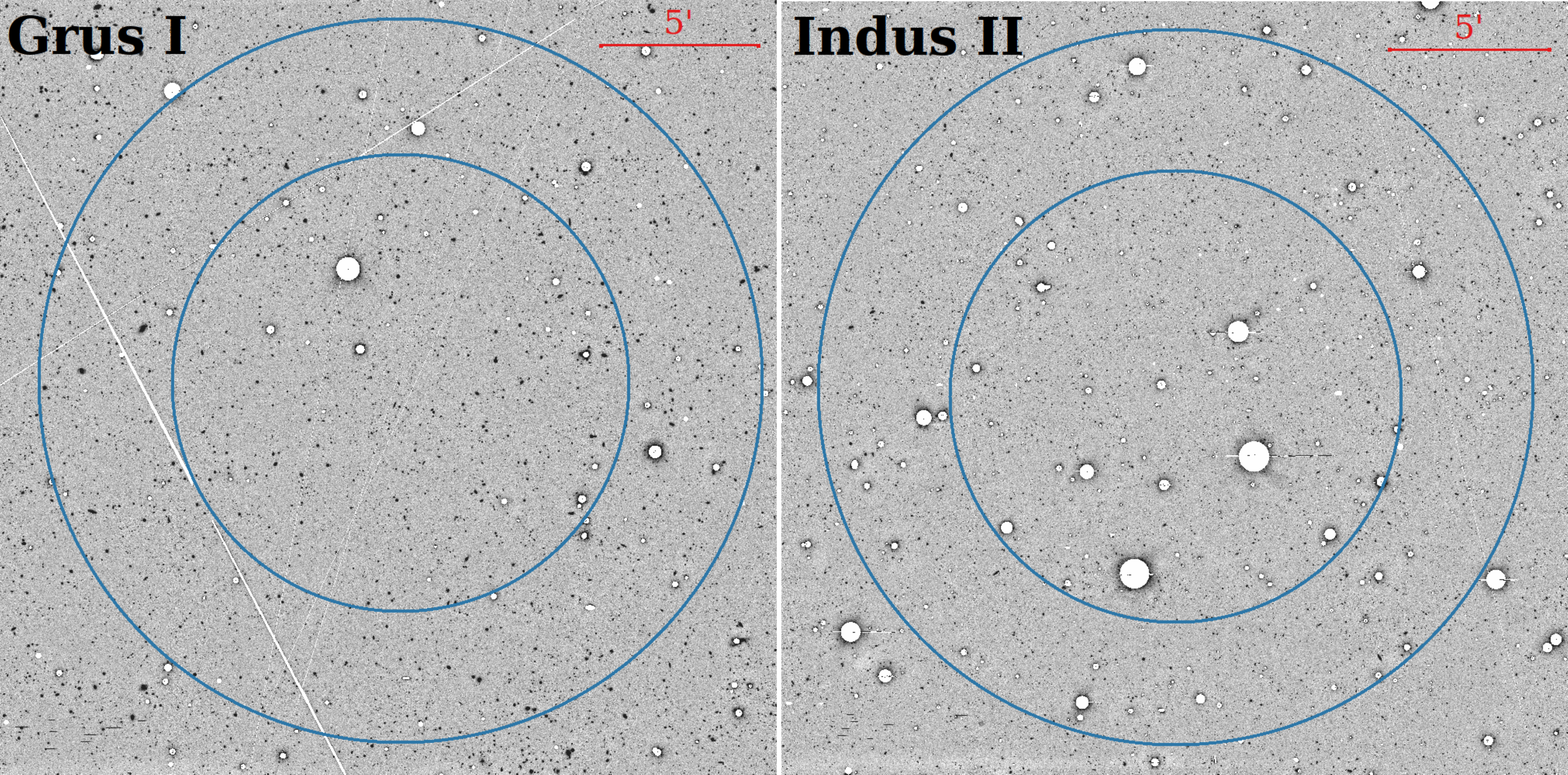}
\end{tabular}   
\caption{Full $~24\arcmin\times24\arcmin$ FOV of Grus I (left panel) and Indus II (right panel).  Shown here are the final SWarped and coadded $r$-band images with masks applied to saturated objects and satellite trails (white marks). The inner blue circles delineate the region of intereset (ROI; see \sref{sec:Anal}) defined in the statistical analysis used to determine final properties of each object. The outer circles mark the outer limit of the area designated as the background region in the statistical analysis.  For both objects, the radii of the circles are $r_{inner}=7\arcmin$ and $7\arcmin \ge r_{background} \ge 12\arcmin$.}
\label{Fig:gruig}
\end{figure*}

\subsection{Data} 
\label{subsec:data}

We observed Indus II and Grus I over four nights in April 2017 with the Megacam instrument \citep{Observatory2015} at the $f/5$ focus of the $6.5$ m Magellan Clay telescope. Megacam is an imager composed of 36 CCDs of $2048 \times 4608$ pixels, creating a square array with a FOV of $\sim 24\arcmin \times 24 \arcmin$ (see Figure \ref{Fig:gruig}). The data were binned $2\times2$ resulting in a pixel scale of $0 \farcs 16$. Observations were dithered such that each image is offset by $+5\arcsec$ in right ascension (RA) and $+13\arcsec$ in declination (Dec.) from the previous one. This reduces the impact of the small gaps between the CCDs. 

The data were reduced using the Megacam pipeline developed at the Harvard-Smithsonian Center for Astrophysics\footnote{This paper uses data products produced by the OIR Telescope Data Center, supported by the Smithsonian Astrophysical Observatory.} \citep{2006ASSL..336..337M}.  This pipeline includes tasks such as bias subtraction, flat fielding, and cosmic ray correction.  In addition, the pipeline derives astrometric solutions using the 2MASS survey \citep{2006AJ....131.1163S}.  The images were then resampled, with a \texttt{lanczos3} interpolation function, and combined with a weighted average using \texttt{SWarp} \citep{Bertin2010}.  This process produced a final, stacked $g$- and $r$-band image for each object. An observing log can be found in Table \ref{Tbl:obs}.  

\begin{deluxetable}{ccccc}[h]
\tablecaption{\label{Tbl:obs}Observing log of Magellan/Megacam observations in the $g$- and $r\text{-bands}$ for Grus I and Indus II.}
\tablecolumns{5}

\tablehead{
Object & UT Date &  Filter & $N\times t_{exp}$ &  Seeing  \\
 & & & (s) & (\arcsec) }
\startdata
 Gru I & 2017 Apr 23 & $g$ & $7\times 300$ & $0.7$ \\ 
  &  2017 Apr 24 & $r$ & $8\times 300$ & $0.9$ \\
 Indus II & 2017 Apr 21 & $g$ & $8\times 300$ & $0.6$ \\
 & 2017 Apr 22 & $r$ & $8\times 300$ & $0. 5$  \\
\enddata
\end{deluxetable}

\begin{deluxetable*}{rcccccccc}[t]
\tablecolumns{9}
\tablecaption{\label{Tbl:phot_sol}The photometric transformations between Megacam instrumental magnitudes and the DES photometric system. The fifth column displays the difference between the transformed Megacam magnitudes and the DES magnitudes. The sixth (sharp range) and seventh (median chi-value) columns detail the morphological cuts made on the instrumental photometry. The last two columns detail the number of stars and magnitude range of the final Megacam$+$DES stellar catalogs after these cuts were applied.}
\tablehead{Object & Filter & $\beta$ & $\alpha$ & std$(\Delta_{\text{mag}})\tablenotemark{$a$}$ & Sharp & Chi & \# of stars & Magnitude range\tablenotemark{$b$}}
\startdata
Grus I & $g$ & 7.554 & $-0.136$ & 0.036  & $(-0.7, 1.2)$  & 1.65 & 6743 & $(15.6, 26.7$)\\
 & $r$ & 7.651 & $-0.027$   & 0.023  & $(-0.5, 0.7)$ & $1.25$ & 6743 & $(15.2, 26.3)$\\
Indus II & $g$  & 7.596 & $-0.167$ & 0.028 & $(-0.5, 0.3)$ & 2.05 & $5520$ & $(15.2, 26.6)$ \\
 & $r$  & 7.657  & $-0.029$   & 0.021 & $(-0.7, 0.2)$ & 4.91 & $5520$ &$(14.8, 26.8)$\\
\enddata
\tablenotetext{$a$}{The median absolute standard deviation of the difference between DES magnitudes and transformed Megacam magnitudes.}
\tablenotemark{$b$}{The faint magnitude limits correspond to $S/N\sim 5$}
\end{deluxetable*}

\subsection{Megacam Photometry} \label{subsec:phot}

Due to the large FOV and number of objects in each image, we used point-spread function (PSF) fitting software to extract the stellar photometry. We used the well-known photometry package, \texttt{DAOPHOT}/\texttt{ALLSTAR} and \texttt{ALLFRAME}, and followed the general guidelines as described in various other papers to determine instrumental magnitudes \citep{1987PASP...99..191S, 1994PASP..106..250S}. 

An accurate PSF model was created from the brightest and most isolated unsaturated stars in the image.  An initial coordinate list and aperture photometry pass of each image was done to find appropriate stars to be used in creating the PSF models. We chose 500 of the brightest stars, evenly distributed over the image, and visually inspected the surrounding areas and radial profiles for saturation, neighbors, bad pixels, and other effects that might affect the measurement of an object. In order to represent stars over the entire FOV, we ensured that the remaining stars were distributed over the entire image and allowed the PSF to vary quadratically. It should be noted that due to the elongation of objects in the Grus I $g$-band image, the fitting radius was set to be slightly larger than the FWHM to better encompass the core of the star. The elongation is along the East-West axis and likely due to tracking issues. 

In order to create a final coordinate list, \texttt{ALLSTAR} was used twice to perform preliminary PSF photometry on the images.  The first run produced a star-subtracted image on which \texttt{ALLSTAR} was run the second time and the stars used in the psf-fit and neighbors were visually inspected. This allows for the detection of fainter objects, located in the PSF wings of brighter objects.  The resultant object list is then input to \texttt{ALLFRAME} to perform a final round of PSF photometry on each filter simultaneously.  In order to convert pixel coordinates from one filter to another, \texttt{DAOMATCH/DAOMASTER} is used to find a linear transformation between the $g$- and $r$-bands for each image. This last step creates a final catalog in each filter that is matched by object ID. It also mitigates the systematic uncertainty created by blended stars being inaccurately measured as one star in some frames.

\subsection{DES Photometry} \label{subsec:des_phot}

We used DES photometry to transform Megacam instrumental magnitudes to DES standard magnitudes and to find magnitudes for the stars saturated in Megacam.  DES is a wide-field survey imaging 5000 deg$^2$ of the southern hemisphere \citep{Abbott2005TheSurvey,2016MNRAS.460.1270D}. DES uses the Dark Energy Camera \citep[DECam;][]{2015AJ....150..150F} positioned at the prime focus of the 4-meter Blanco telescope at the Cerro Tololo Inter-American Observatory (CTIO) in Chile.  DES data are reduced by the DES Data Management (DESDM) pipeline; in which they are detrended, astrometrically calibrated to 2MASS, and coadded into image tiles \citep{2018PASP..130g4501M}.  Detrending includes standard bias subtraction, CCD cross talk, flat fielding, and non-linearity, pupil, fringe, and illumination corrections. Object detection, photometric, and morphological measurements were performed with \texttt{SExtractor} followed by multi-epoch and single-object fitting \citep[SOF;][]{2018PhRvD..98d3526A}.  

The DES catalogs used in this work were created from the DES Y3 GOLD (v2.0) catalog with the selection flags \texttt{FLAGS\_GOLD}$=0$ and \texttt{EXTENDED\_CLASS\_MASH\_SOF}$\leq 2$ in order to ensure we have a complete stellar sample with minimal contaminants. The \texttt{FLAGS\_GOLD} selection applies a bitmask for objects that have known photometric issues and artifacts. The \texttt{EXTENDED\_CLASS\_MASH\_SOF} is similar to the extended classification variables defined in Equations 1, 2, and 3 in \citet{2018ApJ...862..114S}, but for the SOF photometry. The variables in these equations classify objects as high-confidence stars, low-confidence stars, and low-confidence galaxies.

% Due to the magnitude depth being explored in this paper, no foreground object masking was performed.  For details regarding masks of bad regions see \S 7.4 in \citet{2018PhRvD..98d3526A}. The DES catalogs incorporate chromatic and zeropoint corrections, as well as star-by-star Galactic extinction \citep{1998ApJ...500..525S,2011ApJ...737..103S}; all work from this point forward is considered de-reddened.  The DES catalogs were matched to the Megacam catalogs to within $0\farcs25$ using the \texttt{astropy} package \texttt{SkyCoord} \footnote{\cite{astropy:2013}}. The resulting matched catalogs used to transform the Megacam instrumental magnitudes to the DES system had 586 stellar objects for Grus I and 1122 stellar objects for Indus II.

\begin{figure}[h]
\centering
% \plottwo{GruIdist.pdf}{IndIIdist.pdf}
\includegraphics[width=.95\linewidth]{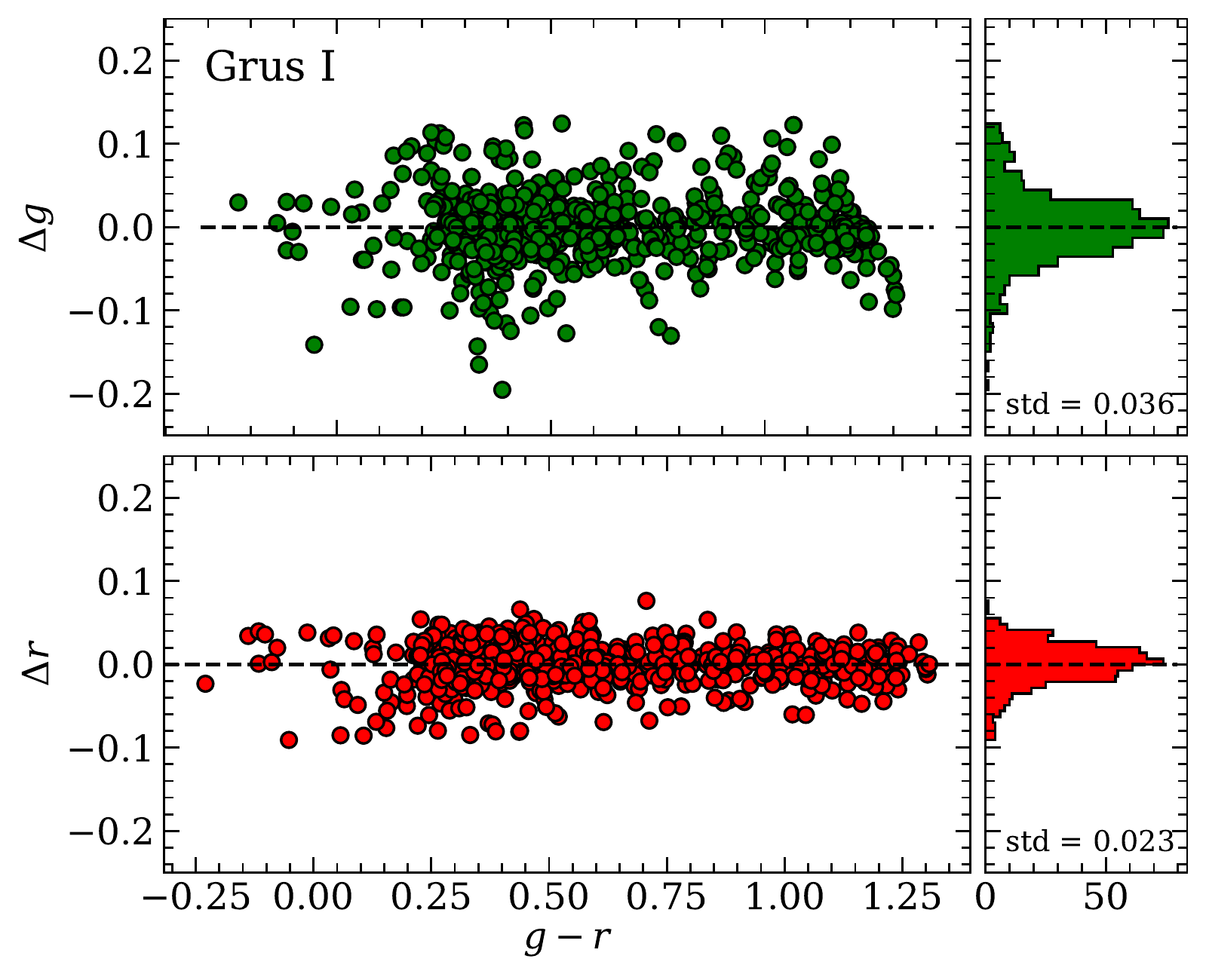}
\includegraphics[width=.95\linewidth]{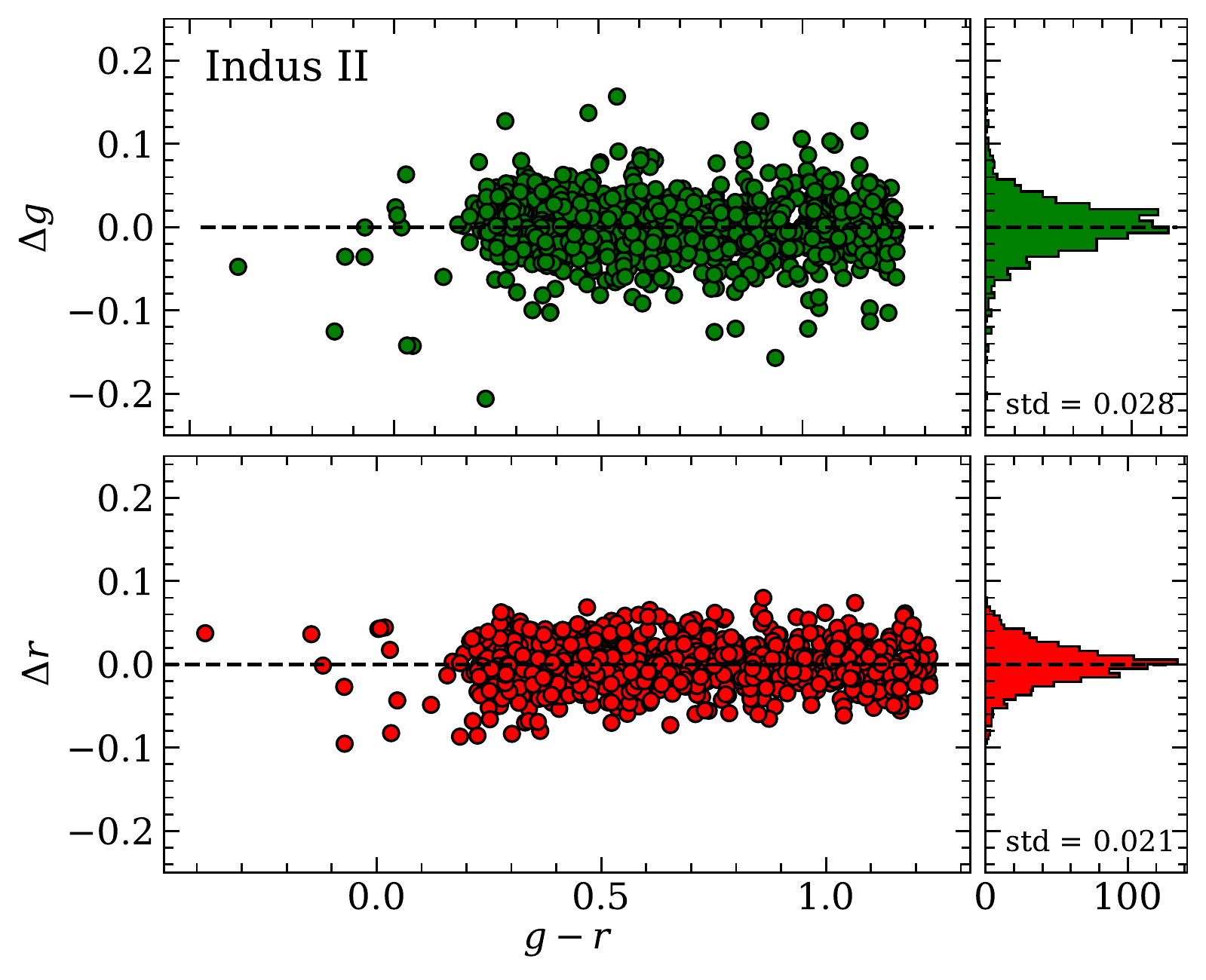}
\caption{Distribution of high confidence stars used in photometric transformation from Megacam instrumental magnitudes to the DES magnitude system. All four panels show $m_{DES} - m_{Megacam}$ vs. $g-r (Megacam)$, where $Megacam$ here are transformed into the DES system. The top two panels show Grus I (586 matched stars) and the bottom two panels show Indus II (1122 matched stars). For both objects, the green points represent $g$-band data and the red points represent $r$-band data.}
\label{Fig:phot_distr}
\end{figure}

\subsection{Transformation from Megacam to DES} \label{subsec:transf}

The matched objects found in the previous section were used to find a transformation between DES magnitudes and Megacam instrumental magnitudes. A color cut of $(g_0-r_0)_{DES} < 1.2$ was applied to remove a clump of M0 and redder stars.  We used only stars having DES photometric errors less than 0.03 mag.  These criteria ensure that a high-quality stellar sample is utilized in finding the magnitude system transformation. 

To perform this transformation, we solve for the coefficients of the following equation using a generalized least squares regression: 
\begin{equation} \label{eq:mags}
M_{DES} = m_{instr} + \beta + \alpha(g_0 - r_0)_{DES},
\end{equation}
where $\beta$ is the zeropoint offset and $\alpha$ is the color coefficient. To find the true distribution of $M_{DES} - m_{instr}$, we run the catalog through a sigma-clipping algorithm based on the median absolute deviation.  Stars that lie outside $3 \sigma$ are clipped until the distribution converges.  Equation \ref{eq:mags} is then applied to all of the instrumental magnitudes found from \texttt{ALLFRAME}. A second-order fit was explored and deemed unnecessary.  The coefficients of this fit are in the third and fourth columns in Table \ref{Tbl:phot_sol} and the difference between transformed Megacam magnitudes and DES magnitudes of stars used to find the transformation can be seen in Figure \ref{Fig:phot_distr}.  

We created the final stellar catalog by applying morphological cuts using the statistics \textit{sharp} and $\chi$ which were determined during the PSF fitting.  \textit{Sharp} can be approximated as \textit{sharp}$^2 \sim \sigma_{obs}^2 - \sigma^2_{PSF}$, where $\sigma_{obs}$ is the observed photometric error and $\sigma_{PSF}$ is the expected photometric error \citep{1987PASP...99..191S}. 

The second statistic, $\chi$, is the ratio of observed pixel-to-pixel scatter over expected scatter, determined from the intrinsic scatter in the PSF models.  Star galaxy separation begins to break down at fainter magnitudes, i.e., $g_0\sim 25.5$ and $r_0\sim 24.75$. The details of these cuts and the magnitude range of the final stellar catalogs can be found in the last four columns in Table \ref{Tbl:phot_sol}.  In addition, the final catalog's brighter magnitudes are supplemented by the DES stellar objects where Megacam saturates at $g_0\sim18$ and $r_0\sim17.5$.  A portion of these catalogs can be seen in Tables \ref{Tbl:Gru_cat} and \ref{Tbl:Ind_cat}.

\begin{figure*}[t]
\centering
\plottwo{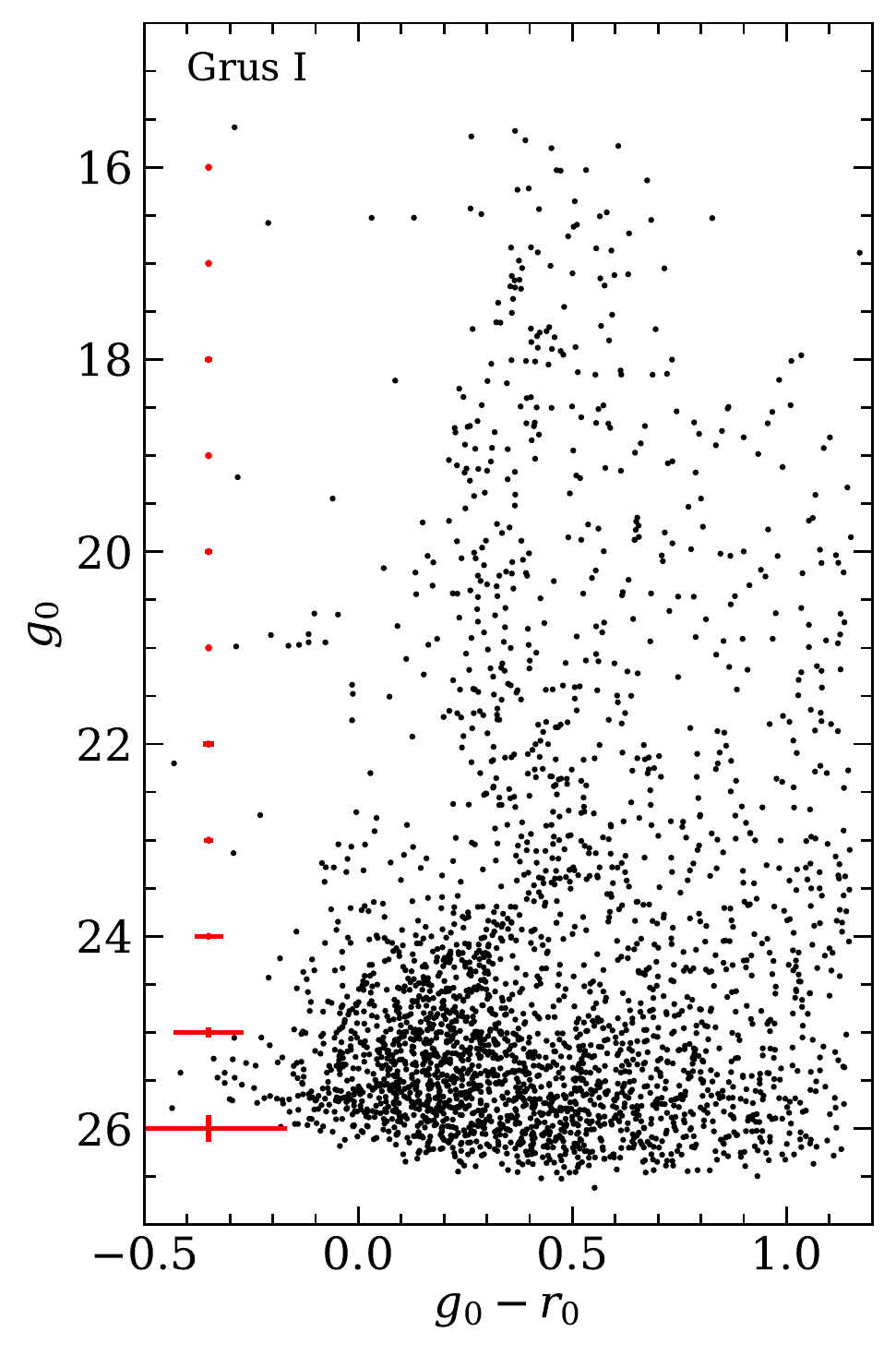}{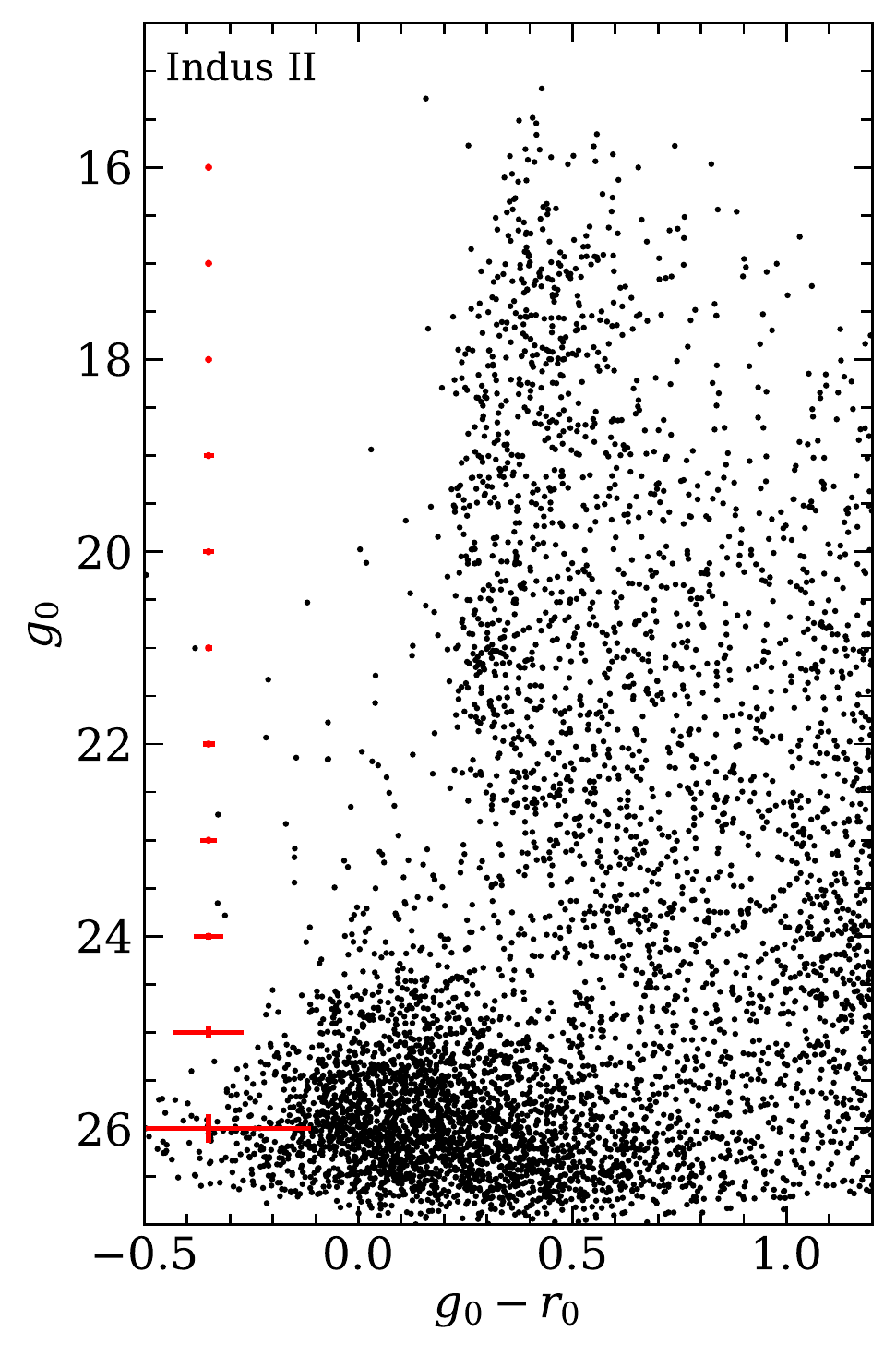}
\caption{$g_0$ vs. $(g_0-r_0)$ CMDs of the full $24\arcmin \times 24\arcmin$ Megacam FOV centered on Grus I (left) and Indus II (right)---created with the final stellar Magellan/Megacam+DES catalog (where objects $g_0\la18$ and $r_0\la 17.5$ are from DES). The error bars represent median photometric uncertainties for one-mag wide bins and are arbitrarily placed in color-space.}
\label{Fig:discoveryCMD}
\end{figure*}

Figure \ref{Fig:discoveryCMD} shows the color-magnitude diagrams (CMDs) created for both Grus I and Indus II using the final calibrated stellar catalog as it was described in this section. The uncertainties show that the photometric signal-to-noise $\sim 10$ to a depth $\sim 3$ magnitudes below that of the discovery papers.

\begin{deluxetable}{ccccccc}[h]
\setlength\tabcolsep{2.9pt}
\tablecolumns{7}
\tablecaption{\label{Tbl:Gru_cat}The final calibrated stellar catalog for Grus I---Sorted by star ID. This table is published in its entirety in machine-readable format. A portion is shown here for guidance regarding its form and content. All magnitudes are in the DES magnitude system.}
\tablehead{Star ID & R.A. & Dec. & $g_{0,DES}$  &  $\sigma_g$  & $r_{0,DES}$  & $\sigma_r$ \\
 & & & (mag) & (mag) & (mag) & (mag)}
\startdata
12543 & 344.134 & $-50.285$ & 25.398 & 0.106 & 24.807 & 0.068 \\
12845 & 344.139 & $-50.284$ & 22.623 & 0.007 & 22.402 & 0.007 \\
13343 & 344.138 & $-50.281$ & 24.075 & 0.029 & 24.044 & 0.036 \\
14597 & 344.103 & $-50.278$ & 24.745 & 0.047 & 24.361 & 0.049 \\ 
14730 & 344.168 & $-50.278$ & 24.956 & 0.070 & 24.902 & 0.076 \\
15406 & 344.187 & $-50.275$ & 17.679 & 0.003 & 17.276 &  0.002 \\
\enddata
\end{deluxetable}

 \begin{deluxetable}{ccccccc}[h]
 \setlength\tabcolsep{2.9pt}
\tablecolumns{7}
\tablecaption{\label{Tbl:Ind_cat}The final calibrated stellar catalog for  Indus II---Sorted by star ID. This table is published in its entirety in machine-readable format. A portion is shown here for guidance regarding its form and content.}
\tablehead{Star ID & R.A. & Dec. & $g_{0,DES}$  &  $\sigma_g$  & $r_{0,DES}$  & $\sigma_r$ \\
 & & & (mag) & (mag) & (mag) & (mag)}
\startdata
17138 & 309.735 & $-46.284$ & 25.845 & 0.125 & 25.532 & 0.090 \\
17539 & 309.705 & $-46.283$ & 25.911 & 0.117 & 25.658 & 0.100 \\
17819 & 309.691 & $-46.282$ & 21.532 & 0.007 & 20.891 & 0.006 \\
18608 & 309.677 & $-46.278$ & 21.129 & 0.005 & 20.183 & 0.010 \\ 
18821 & 309.764 & $-46.278$ & 25.005 & 0.064 & 24.897 & 0.052 \\
19208 & 309.689 & $-46.276$ & 21.143 & 0.006 & 20.198 &  0.010 
\enddata
\end{deluxetable}

\section{Methods}\label{sec:Anal}

We utilize the Ultra-faint Galaxy Likelihood (\texttt{UGaLi})\footnote{\url{https://github.com/DarkEnergySurvey/ugali}} toolkit to determine structural parameters and the best-fitting isochrones for Grus I and Indus II.
Here we review the aspects of \texttt{UGaLi} that are important for our analysis, and refer to~\citet{Bechtol2015} and the appendix of \citet{drlica-wagner2019} for a more detailed description.  

Our data sample consists of the magnitude and the error on the magnitude in two filters, ${\cal D}_{c,i} = \{g_i, \sigma_{g_i}, r_i, \sigma_{r_i\}}$, and the spatial positions of the stars ${\cal D}_{s,i} = \{ \alpha_i, \delta_i \}$. We define the probability distribution for the structural parameters as $u_s$, and the probability distribution for the parameters of the isochrone as $u_c$. 
The total probability distribution function (PDF) for the data ${\cal D}_i = \{ {\cal D}_{s,i}, {\cal D}_{c,i} \}$ given the model parameters $\bm{\theta}$ is then \begin{equation}\label{Eq:pdf}
    u(\mathcal{D}_i|\bm{\theta}) = u_s(\mathcal{D}_{s,i}|\bm{\theta}_s)\times u_c(\mathcal{D}_{c,i}|\bm{\theta}_c), 
\end{equation}
This probability distribution is defined such that the integral of it over the entire spatial and magnitude domain is unity. 

For the structural properties, $u_s$, we assume an elliptical Plummer model, with a projected density distribution~\citep{doi:10.1093/mnras/71.5.460,Martin2008},
\begin{equation}\label{Eq:prof}
\Sigma (R) \propto \left[ 1+\left(\frac{R_i}{R_p}\right)^2\right]^{-2}. 
\end{equation}
Here $R_i$ is the elliptical radius coordinate from the center of the galaxy, and $R_p$ is the Plummer-scale radius (equivalent to the 2D azimuthally averaged half-light radius, $r_h = a_h\sqrt{1-\epsilon}$). There are five model parameters that describe the Plummer profile: the centroid coordinates ($\alpha_0$, $\delta_0$,), the semi-major half-light radius ($a_h$), the ellipticity ($\epsilon$), and position angle ($\phi$). The density distribution is further related to spatial position by 

 \begin{equation}
 \begin{split}
     R_i  =  \Bigg\{\left[\frac{1}{1-\epsilon}\left(X_i\cos{\phi}-Y_i\sin{\phi}\right)\right]^2 & \\
      - (X_i\sin{\phi} + Y_i\cos{\phi})^2\Bigg\}^\frac{1}{2} &
\end{split}
 \end{equation}
 
 and spatial position is related to the object centroid by
 \begin{equation}
     X_i - X_0 = (\alpha_i - \alpha_0)\cos(\delta_0)
\end{equation}
and
\begin{equation}
     Y_i - Y_0 = \delta_i - \delta_0.
 \end{equation}

For the isochrone properties, $u_c$, we calculate the PDF by binning the color-magnitude information over a grid of isochrones that are weighted by a Chabrier IMF \citep{chabrier2003} and have a fixed solar alpha abundance. These isochrones are described in terms of the distance modulus ($m-M$), the age of the stellar population ($\tau$), and the metallicity ($Z)$. All metallicities are reported as [Fe/H] $=\log_{10}\left(\frac{Z}{Z_\sun}\right)$, with $Z_\sun=0.0152$. 

The grid of PARSEC isochrones are representative of old metal-poor stellar populations, i.e., $0.0001 < Z < 0.001$, 1 Gyr $< \tau <$ 13.5 Gyr, and $16.0< m-M < 25.0$ to fit the CMD properties of each object \citep{2012bressanIso}.   We check that our results do not depend on the isochrone model by comparing to \citet{2016DotterIso} and find that they are insensitive to this specific assumption.

With the above model, we can define the Poisson log-likelihood 

\begin{equation}\label{Eq:log}
    \log \mathcal{L} = -\lambda N_s - \sum_i^{stars}
    \log (1-p_i),
\end{equation}

\noindent where $\lambda$, the stellar richness, is a normalization parameter representative of the total number of member stars with $M_*>0.1M_\sun$ in the satellite, $N_s$ is the fraction of observable satellite member stars, and $p_i$ is the probability that a star is a member of the satellite.  

Because we choose to normalize the signal PDF to unity, we can interpret $\lambda$ as the total number of stars in the satellite (observed $+$ unobserved). The membership probability is given by 

\begin{equation}\label{Eq:probs}
    p_i = \frac{\lambda u_i}{\lambda u_i + b_i},
\end{equation}

\noindent where $b_i$ is the background density function \citep[for more details see Appendix C in][]{selection2019}.  We take the background density function to be independent of spatial position in our region of interest (ROI). 

The empirical background density function, $b_i$, is determined from an annulus ($7\farcm2 < r < 12\arcmin$) surrounding our target ROI ($r < 7\farcm2$). We require the ROI to be $\gtrsim 2\times r_h$. This is the maximum ROI that still allows for the background annulus to contain $\sim 3\times r_h$ \citep{Martin2008}, where $r_h$ is from \cite{Koposov2015, TheDESCollaboration2015}. Figure \ref{Fig:gruig} depicts these regions as blue circles.  Any non-stellar objects that still contaminate the data at greater magnitudes are expected to do so equally over the entire FOV and therefore averaged within $b_i$. 

With $\lambda$ allowed to vary and $b_i$ held fixed, we simultaneously explore the whole parameter space with flat priors for all parameters except $r_h$ (an inverse prior). With \texttt{UGaLi}, we run an MCMC chain with 100 walkers, 12000 steps, and 1000 burn-in \citep{2013PASP..125..306F}. Absolute $V$-band magnitude is determined following the prescription of \citet{Martin2008}

\section{Results \& Discussion} \label{sec:Res}

\subsection{Grus I} \label{subsec:gru}

\begin{figure*}[p]
\centering
\plotone{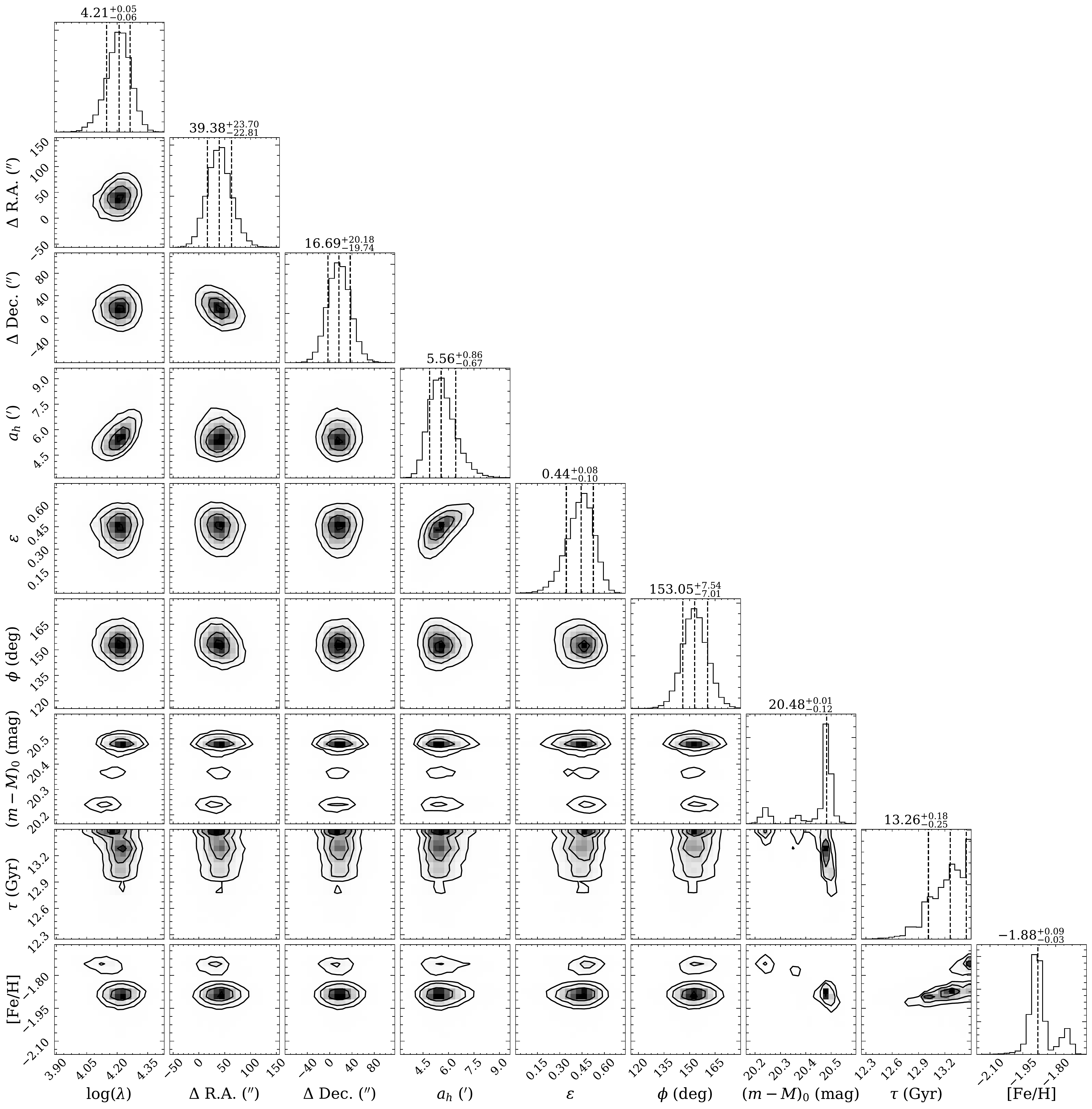}
\caption{Posterior probability distributions for the structural and isochrone parameters of Grus I obtained from an elliptical Plummer model and grid of PARSEC isochrones. The parameters explored were (from left to right): stellar richness ($\lambda$), $\Delta$ R.A. \& $\Delta$ Dec.(these are the shift from the centroid found in \citet{Koposov2015}), semi-major half-light radius ($a_h$), ellipticity ($\epsilon$), position angle ($\phi$), distance modulus ($(m-M)_0$), age ($\tau$), and metallicity ([Fe/H]). Dashed lines in the 1D histograms indicate 16th, 50th, and 84th quantiles of the median peak likelihood. We have excluded these quantiles from $(m-M)_0$ and [Fe/H] due to their bimodality.}
\label{Fig:gru_corner}
\end{figure*}

\begin{figure}[t]
\centering
\plotone{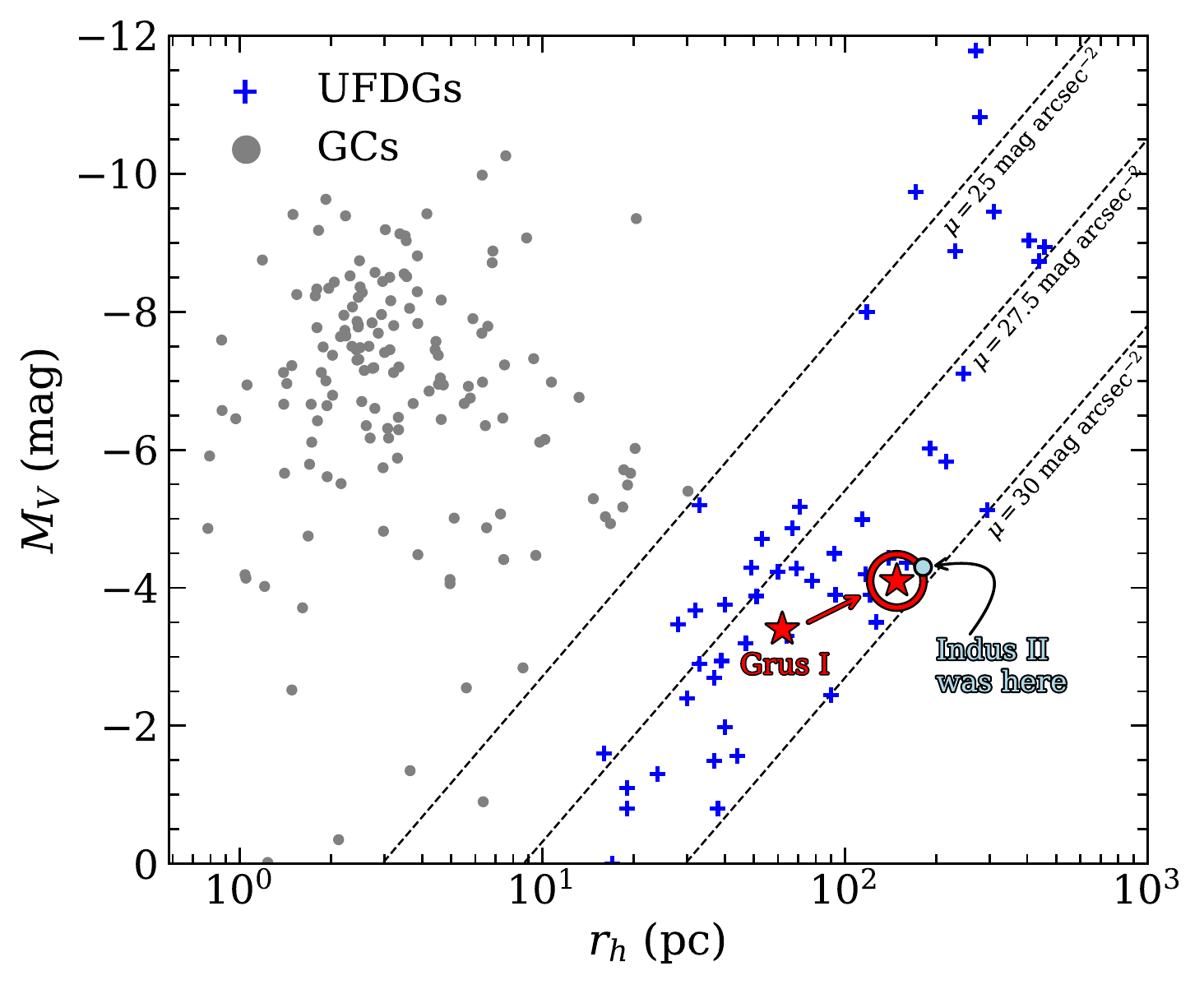}
\caption{Shown as a red star are the position of Grus I with our newly derived properties (circled in red) and its original location in the size-luminosity plane. The original location of Indus II is shown as a light blue circle. The MW globular clusters are in grey points and the rest of the MW UFDs are depicted with blue crosses.}
\label{Fig:sizelum}
\end{figure}

Column 1 of Table \ref{Tbl:MLparams} lists the parameters obtained from the median peak likelihood of the posterior distributions (see Figure \ref{Fig:gru_corner}). With our improved parameters for Grus I, we find it to be consistent with an extended ultra-faint dwarf galaxy that resides at the faint edge ($\mu \sim 30$ mag) of the galaxy locus in the size-luminosity plane. Figure \ref{Fig:sizelum} shows the relationship between GCs and UFDs in this parameter space and also shows the updated location of Grus I. Table \ref{Tbl:MLparams} lists the parameters that were derived in previous works (both Grus I and Indus II are represented).  

With a $a_h\sim 202$ pc and $M_V\sim -4.1$, Grus I is both larger and brighter than estimates from previous works \citep{Koposov2015}.  In addition, according to our results ([Fe/H] $\sim -1.9$), it is among the more metal-rich UFDs found to date \citep{simon2019faintestdwarfs}. Spectroscopic studies by \citet{Walker2016, 2019ApJ...870...83J} find the brightest potential member stars are very metal poor (e.g., [Fe/H]~$\sim-2.3$), whereas photometric studies (including this one) find it to be less metal poor, i.e., [Fe/H]~$\lesssim-2$ \citep[][]{Koposov2015, 2018arXiv180902259J}. A larger spectroscopic sample is required to confirm the metallicity of this object. 

This discrepancy between spectroscopic and photometric metallicities has been seen in previous studies \citep[see Section 4.3 in][]{crater2}. In that case, it was considered more likely that the spectroscopic results were likely systematically metal-poor.  There was very good agreement with the isochrone calculated with the photometric metallicity and probable member stars.  

\begin{deluxetable*}{lcccccccc}[t]
\centering
\setlength\tabcolsep{2.9pt}
\tablecolumns{9}
\tablecaption{\label{Tbl:MLparams}Photometric and spectroscopic parameters of Grus I and Indus II found in the literature prior to this work. The columns for Grus I, in order, are from this work, \cite{Koposov2015, Walker2016, 2018arXiv180902259J, Munoz2018, 2019ApJ...870...83J, martinez19}. Column 8 describes Indus II  as found in the discovery paper, \cite{TheDESCollaboration2015}.}
\tablehead{ & \multicolumn{7}{c}{Grus I} &  Indus II \\ & This Work \\
&  (1) & (2) & (3) & (4) & (5) & (6) & (7) & (8)}
\startdata
$\alpha_{2000}$ (deg) & $344.166^{+0.007}_{-0.006}$ & 344.1765 & \nodata & $344.1700$ & 344.1797 & \nodata & \nodata & 309.76\\
$\delta_{2000}$ (deg) & $-50.168_{-0.005}^{+0.006}$ & $-50.1633$ & \nodata & $-50.1641$ &  $-50.1800$ & \nodata &  \nodata &$-46.16$ \\
t-value ($\sigma$) & $21.3$ & 10.1 & \nodata & \nodata & \nodata & \nodata & \nodata & $\sim5.7$ \\
$M_V$ (mag) & $-4.1\pm0.3$ & $-3.4\pm0.3$ & \nodata & \nodata & \nodata & \nodata & \nodata & $-4.3\pm0.19$ \\
$D_\sun$ (kpc) & $125_{-12}^{+6}$ & 120 & \nodata  & $115\pm6$ & \nodata & \nodata &  \nodata &$214\pm16$ \\
$r_h$ (arcmin) & $4.16_{-0.74}^{+0.54}$ & $1.77^{+0.85}_{-0.39}$\tablenotemark{$b$} & \nodata & \nodata & $0.81\pm0.66\tablenotemark{$b$}$ & \nodata &  \nodata &$2.9^{1.1}_{1.0}$\tablenotemark{$b$}  \\
$r_h$ (pc) & $151^{+21}_{-31}$ & $62^{+29.8}_{-13.6}$ & \nodata & \nodata & $28.3\pm23.0$ & \nodata &  \nodata &$181\pm67$ \\
$\epsilon$ & $0.44_{-0.10}^{+0.08}$ & $0.41^{+0.20}_{-0.28}$ & \nodata & \nodata & $0.45\pm0.30$ & \nodata & \nodata & $<0.4$ \\
$\phi$ (deg) & $153_{-7.0}^{+8.0}$ & $4\pm 60$ & \nodata & \nodata & $23\pm18$ & \nodata & \nodata & \nodata \\
$m-M$ (mag) & $20.48_{-0.22}^{+0.11}$\tablenotemark{$a$} & 20.4 & \nodata & $20.30\pm0.11$ & \nodata & \nodata & $20.51\pm 0.10$\tablenotemark{$c$} & \nodata \\
$\tau$ (Gyr) & $13.26^{+0.18}_{-0.25}$ & \nodata & \nodata & $14.0^{+1.0}_{-1.0}$ & \nodata & \nodata & \nodata & \nodata \\
$[Fe/H]$ (dex) & $-1.88^{+0.09}_{-0.03}$ & \nodata & $-1.42^{+0.55}_{-0.42}$ & $-2.5^{+0.3}_{-0.3}$ & $-2.5\pm0.3$ & $-2.57,-2.50$ &  \nodata &\nodata  \\
$\sigma_{[Fe/H]}$ (dex) & \nodata & \nodata & $<0.9$ & \nodata & \nodata & \nodata & \nodata & \nodata  \\
$\left<v_{LOS}\right>$ (km s$^{-1}$) & \nodata & \nodata & $-140.5^{+2.4}_{-1.6}$ & \nodata & \nodata & \nodata &  \nodata &\nodata \\
$\sigma_{v_{LOS}}$ (km s$^{-1}$) & \nodata & \nodata & $<9.8$ & \nodata & \nodata & \nodata  &  \nodata &\nodata \\
$M_\sun / L_{V,\sun}$ & \nodata & \nodata & $<2645$ & \nodata & \nodata & \nodata  & \nodata & \nodata 
\enddata
\tablenotetext{a}{A systematic uncertainty of 0.1 mag was added to account for the difference between best-fits for \citet{2012bressanIso} and \citet{2016DotterIso} isochrones.}
\tablenotetext{b}{Semi-major halflight radii converted from azimuthally-averaged radii with $\sqrt{1-\epsilon}$ factor.}
\tablenotetext{c}{This distance measurement is based on two RR Lyrae stars found in Grus I.}
\end{deluxetable*}

\begin{figure*}[t]
\centering
\begin{tabular}{c}
   \includegraphics[width=1\linewidth]{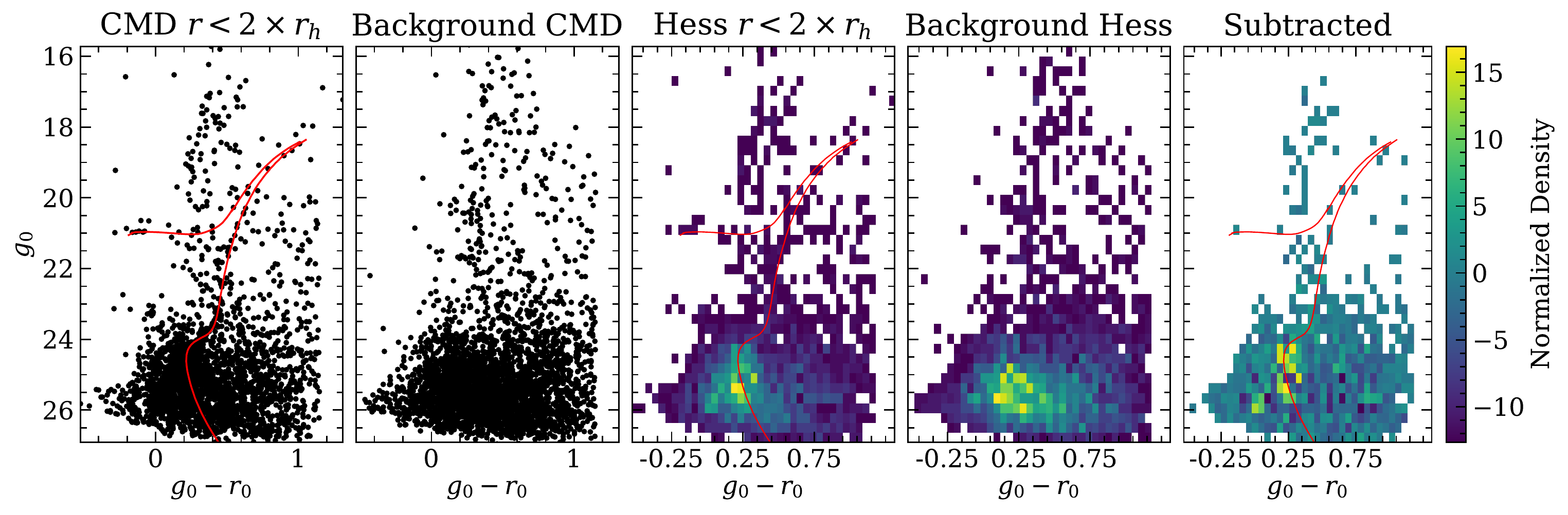} \\
   \includegraphics[width=1\linewidth]{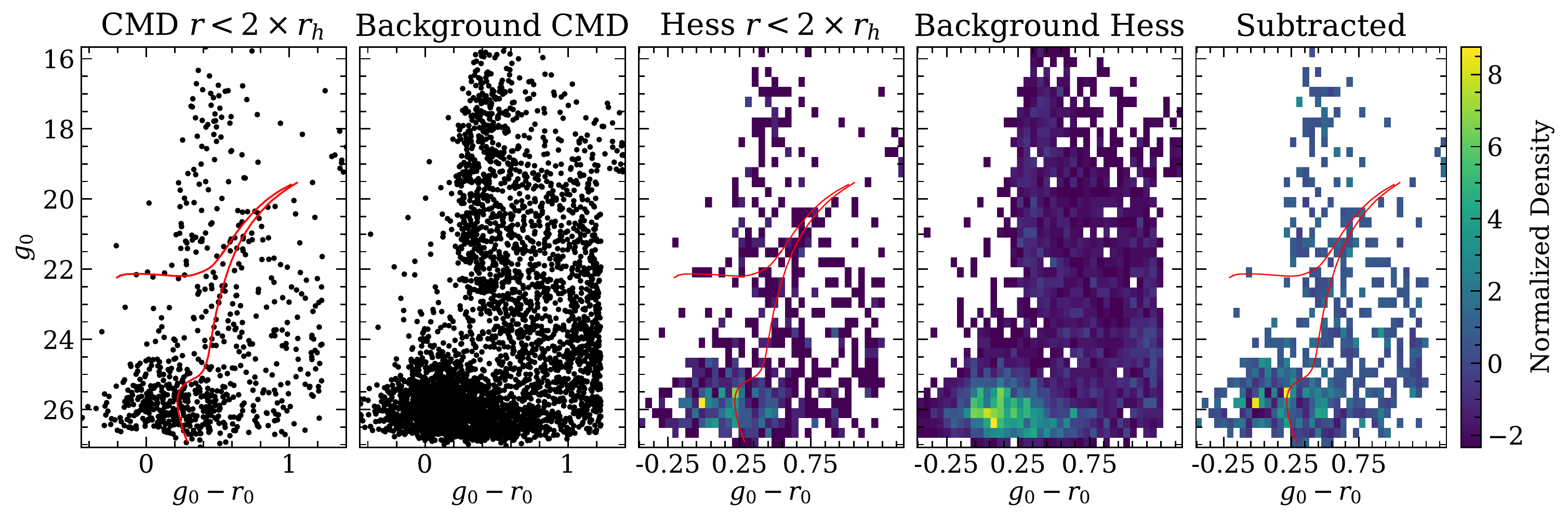}
\end{tabular}   
\caption{The top row of panels is for Grus I CMD and the bottom row is for Indus II CMD. In both cases, the 1st panel shows an ROI of $2\times r_h$ centered on the object, where Grus I uses properties found in this study (see Table \ref{Tbl:MLparams}) and Indus II uses the discovery properties from \citet{TheDESCollaboration2015}. The 2nd panel is a comparison CMD made from stars ~$>2\times r_h$ away from the ROI. The 3rd panel is a Hess diagram showing the density of the background stars seen in the 2nd panel.  They have been scaled to match the same area as the ROIs. The 4th panel is the Hess diagram of the stars within $2\times r_h$ as seen in the 1st panel. The 5th panel is the Hess difference of the 4th and 3rd panels.}
\label{Fig:hess}
\end{figure*}

Figure \ref{Fig:hess} shows the CMD of the stars within $2\times r_h$ and the CMD of the background (see the 1st and 2nd panel, respectively).  The last three panels of Figure \ref{Fig:hess} are Hess diagrams of the background stellar density, the stellar density within $2\times r_h$, and the difference of the two (see panels 3, 4, and 5, respectively). Overlaid in the 1st, 4th, and 5th panels is a PARSEC isochrone representative of an old, metal-poor population with $\tau=13.3$ Gyr and [Fe/H]~$=-1.9$. This isochrone agrees with the best-fit properties of Grus I inferred from the maximum likelihood distribution. 

For Grus I the background-subtracted Hess diagram shown in the 5th panel of Figure \ref{Fig:hess}, clearly illuminates MSTO, MS, and sub-giant branch features that are well-represented by the inferred properties. Less obvious, but still well-populated, the isochrone clearly delineates a HB and RGB population. It should be noted that some potential members can still be seen in the second top-row panel (background CMD) of Figure \ref{Fig:hess} due to Grus I's large extent.

\begin{figure*}[t]
\centering
\plotone{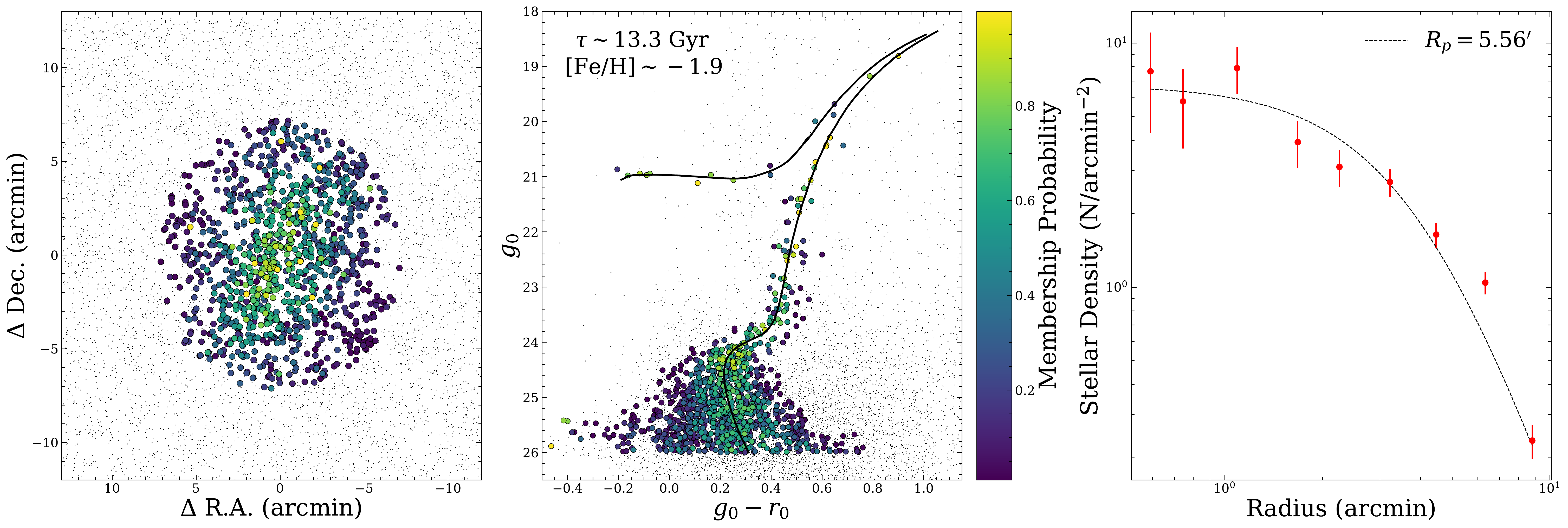}
\caption{Right panel: Spatial distribution of stars in Grus I that have high membership probability. Middle Panel: Color-magnitude diagram of the same stars with high membership probabilty. The black line is the isochrone best described by our newly derived parameters in Table \ref{Tbl:MLparams}. Gray points in both panels are stars with less than 5\% membership probability.Right panel: The stellar density profile of Grus I where the data is shown in red in elliptical bins of equal number and weighted by associated membership probabilities. The black dashed line shows the theoretical two-dimensional Plummer profile created with a Plummer-scale radius equal to $a_h=5.6\arcmin$. }
\label{Fig:grui_probs}
\end{figure*}

In the left two panels of Figure \ref{Fig:grui_probs}, we show the distribution of the \texttt{UGaLi} membership probabilities in sky coordinates and color-magnitude space.  These membership probabilities were determined as described in \sref{sec:Anal}. The three panels in this figure show that our inferred parameters describing the stellar population and morphology of Grus I are consistent with a theoretical Plummer profile.

These probabilities were further used to create a binned and weighted density profile as seen in the far right panel of Figure \ref{Fig:grui_probs}. There are an equal number of stars in each bin.  It can be seen that the binned data fits well over the Plummer model profile shown as the dashed line.  

The posterior distributions and maximum-likelihood peak values are shown in Figure \ref{Fig:gru_corner}. While some of the properties shown in Figure \ref{Fig:gru_corner} agree with previous works (see Table \ref{Tbl:MLparams}) within the uncertainties (e.g., centroid coordinates, ellipticity, distance modulus), others have shifted slightly in this work (i.e., $r_h$), changing some of the derived properties. 

\begin{figure}[t]
\centering
\plotone{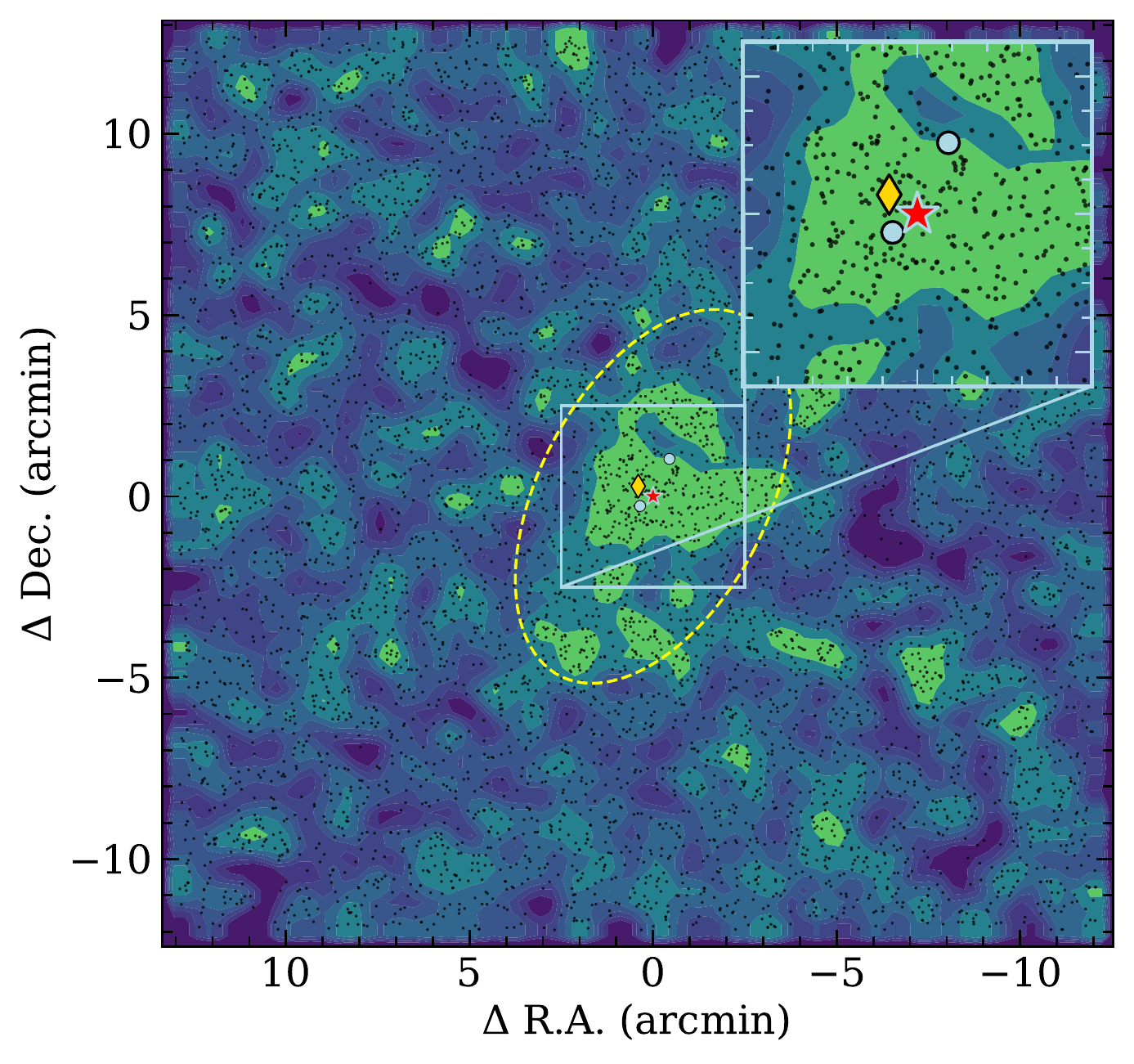}
\caption{The $3-\sigma$ iso-density contours of Grus I with our redetermined centroid shown as a red star.  The blue circles represent the location of the two overdensities mentioned in \citet{2018arXiv180902259J} and the gold diamond is the original centroid found in \citet{Koposov2015}. The yellow dashed ellipse indicates the Plummer halflight radius found in this work.}
\label{Fig:gruiProf}
\end{figure}

\citet{2018arXiv180902259J} find two small overdensities at $[(\alpha-\alpha_0),(\delta-\delta_0)]\approx [+0.2,-0.5]$ (arcmin) and $[(\alpha-\alpha_0), (\delta-\delta_0)]\approx [-0.6,+0.8]$ (arcmin)--with extents of $22\times25$ pc and $13\times28$ pc, respectively. It is interesting to note that we do not find obvious evidence of the two slight overdensities or diffuse centroid found in \citet{2018arXiv180902259J}. Our centroid shift does not seem to be significant or dependent on any lack of dense central overdensity as can be seen in Figure \ref{Fig:gruiProf}.  The dashed yellow line in this figure indicates the halflight radius created with our inferred parameters. 

The $r_h$ (4\farcm16) found in this work is larger than previous works by more than a factor of 2. Our larger FOV (see Figure \ref{Fig:gruig}) allows us to more accurately constrain the local background contamination and is likely the reason for the change in extent.  Additionally, we find Grus I to be about one magnitude brighter ($M_V \sim -4.1$ mag) than previously thought \citep{Koposov2015}, while the distance to the object is in agreement with the recently updated distance determination based on RR Lyrae stars \citep{martinez19}.

It should be noted that Jhelum, a nearby stellar stream \citep[$D_\sun\sim13$ kpc, $m-M\sim 15.6$;][]{2018ApJ...862..114S}, potentially contaminates the FOV. In order to test this, we cut potential stream member stars from our catalog and performed the same analysis on the new catalog. These potential members were chosen based on Jhelum's spatial footprint and location in color-magnitude space. The width of the area in color-magnitude space was chosen to account for our photometric uncertainties. The distance of the Jhelum stream ($D_\sun\sim13$ kpc) compared to how far Grus I is precludes any physical association between the two. The results from this analysis were the similar within uncertainties. Therefore, we determined that the presence of the stream does not significantly affect our analysis.

\subsection{Indus II} \label{subsec:indii}

MCMC chains run on this object fail to converge and no membership probabilities are calculated. The resulting isochrone parameters from the uncoverged chains are indicative of a young stellar population, which is inconsistent with UFDs or globular clusters.

In all but the background panels of Figure \ref{Fig:hess}, the UFD representative old, metal-poor isochrone delineates the HB but fails to match with any other CMD feature. The derived isochrone fails to match the BHB stars that its candidacy hinged on originally. This indicates Indus II is likely a false-positive, i.e., neither a dwarf galaxy nor a globular cluster.  

Since we have found that Indus II is neither a real galaxy nor a cluster, it goes against convention to use Indus II as its designation.  We prefer to use the designation \indii~from now on.

\section{Conclusions and Final Remarks}\label{sec:disc}

We confirm the status of Grus I as a likely dwarf galaxy with the results of an MCMC algorithm and fit to a Plummer density model with deep Magellan/Megacam follow-up photometry of the objects Grus I and  \indii. This photometry reaches $\sim 2-3$ magnitudes deeper than the discovery data, allowing us to derive i) improved distance, which is in agreement with the distance obtained using RR Lyrae distance indicators \citep{martinez19}, ii) luminosity, ~1 mag brighter than \citet{Koposov2015}, and iii) structural parameters, particularly finding that the $r_h$ is two times larger. We find that \indii~ is a false positive that was flagged due to a chance projection of an overdensity of stars.

Grus I is an extended ($r_h\approx 4\farcm16$), elliptical ($\epsilon=0.44$) dwarf galaxy with a distance of $125$ kpc.  Like other dwarfs, Grus I has an old single, stellar population ($13.3$ Gyr) with low metallicity ([Fe/H]~$=-1.9$).  Its luminosity ($M_V\approx-4.1$) and azimuthal half-light radius ($151$ pc) place it at the lower edge of the dwarf locus in the size-luminosity plane (see Figure \ref{Fig:sizelum}).

Our analysis complements previous studies with a larger FOV and deeper photometry, allowing us to confirm the suggestion of \citet{Walker2016} that Grus I was likely larger in extent\footnote{This larger extent implies a lower-density dark matter halo\citep{Wolf2010}}. We also find that Grus I is slightly less metal-poor than most UFDs, with [Fe/H]$\lesssim-1.9$ although not as metal-rich as suggested in \cite{Walker2016}. 

In this work, we reach the necessary FOV ($24\arcmin\times 24\arcmin$) which allows us to improve upon and find new structural parameters of Grus I. \citet{Martin2008} determined that a FOV three times the half-light radius is necessary to accurately constrain the structural properties of UFDs. Therefore, with $r_h\sim4\farcm16\arcmin$, our $24\arcmin\times 24\arcmin$ FOV is just large enough to derive accurate structural parameters.

Previous spectroscopic studies find mixed results with respect to the average metallicity of this object.  There are two well-measured, brighter member stars that are consistent with old, metal-poor UFDs ([Fe/H]$\sim-2.5$), but \cite{Walker2016} found five faint stars that suggested a metallicity of [Fe/H]$\sim-1.4$. It is possible that these fainter stars have a systematic uncertainty or bias that cause Grus I to appear more metal-rich than it is.

We analyze \indii~ with background subtracted Hess diagrams and \texttt{UGaLi} and find that the distribution of stars does not correlate with any isochrone or Plummer model. A chance alignment of possible BHB stars contributed to the original detection of \indii~ as a candidate satellite.  Constraining power for this overdensity comes from a set of BHB stars.  This feature can be seen in our dataset as well, but does not match a corresponding MSTO that is consistent with a UFD.  We conclude that  \indii~ is not consistent with either a dwarf galaxy or a globular cluster.

Ongoing follow-up studies of UFDs will continue to have important implications for our understanding of near-field cosmology.  As the most dark matter dominated objects and the only resolved examples of these old, relatively pristine stellar populations, it is important that their nature is well understood. Understanding their nature contributes to the characterization of the satellite population of the MW, leading to more accurate inferences about galaxy formation physics and the nature of dark matter. Studies with deep and wide follow-up photometry, such as this one, are useful to help characterize these faint objects.

\section*{Acknowledgements}
This paper has gone through internal review by the DES collaboration. The authors thank Taylor A. Hutchison, Jonathan H. Cohn, and Peter S. Ferguson for their insightful conversation and support throughout this analysis. SAC acknowledges support from the Texas A\&M University and the George P. and Cynthia Woods Institute for Fundamental Physics and Astronomy. ABP acknowledges support from NSF grant AST-1813881. JDS acknowledges support from NSF grant AST-1412792. This research made use of Astropy,\footnote{http://www.astropy.org} a community-developed core Python package for Astronomy \citep{astropy:2013, astropy:2018}.
This research made extensive use of \href{https://arxiv.org/}{arXiv.org} and NASA's Astrophysics Data System for bibliographic information.

Funding for the DES Projects has been provided by the U.S. Department of Energy, the U.S. National Science Foundation, the Ministry of Science and Education of Spain, 
the Science and Technology Facilities Council of the United Kingdom, the Higher Education Funding Council for England, the National Center for Supercomputing 
Applications at the University of Illinois at Urbana-Champaign, the Kavli Institute of Cosmological Physics at the University of Chicago, 
the Center for Cosmology and Astro-Particle Physics at the Ohio State University,
the Mitchell Institute for Fundamental Physics and Astronomy at Texas A\&M University, Financiadora de Estudos e Projetos, 
Funda{\c c}{\~a}o Carlos Chagas Filho de Amparo {\`a} Pesquisa do Estado do Rio de Janeiro, Conselho Nacional de Desenvolvimento Cient{\'i}fico e Tecnol{\'o}gico and 
the Minist{\'e}rio da Ci{\^e}ncia, Tecnologia e Inova{\c c}{\~a}o, the Deutsche Forschungsgemeinschaft and the Collaborating Institutions in the Dark Energy Survey. 

The Collaborating Institutions are Argonne National Laboratory, the University of California at Santa Cruz, the University of Cambridge, Centro de Investigaciones Energ{\'e}ticas, 
Medioambientales y Tecnol{\'o}gicas-Madrid, the University of Chicago, University College London, the DES-Brazil Consortium, the University of Edinburgh, 
the Eidgen{\"o}ssische Technische Hochschule (ETH) Z{\"u}rich, 
Fermi National Accelerator Laboratory, the University of Illinois at Urbana-Champaign, the Institut de Ci{\`e}ncies de l'Espai (IEEC/CSIC), 
the Institut de F{\'i}sica d'Altes Energies, Lawrence Berkeley National Laboratory, the Ludwig-Maximilians Universit{\"a}t M{\"u}nchen and the associated Excellence Cluster Universe, 
the University of Michigan, the National Optical Astronomy Observatory, the University of Nottingham, The Ohio State University, the University of Pennsylvania, the University of Portsmouth, 
SLAC National Accelerator Laboratory, Stanford University, the University of Sussex, Texas A\&M University, and the OzDES Membership Consortium.

Based in part on observations at Cerro Tololo Inter-American Observatory, National Optical Astronomy Observatory, which is operated by the Association of 
Universities for Research in Astronomy (AURA) under a cooperative agreement with the National Science Foundation.

The DES data management system is supported by the National Science Foundation under Grant Numbers AST-1138766 and AST-1536171.
The DES participants from Spanish institutions are partially supported by MINECO under grants AYA2015-71825, ESP2015-66861, FPA2015-68048, SEV-2016-0588, SEV-2016-0597, and MDM-2015-0509, 
some of which include ERDF funds from the European Union. IFAE is partially funded by the CERCA program of the Generalitat de Catalunya.
Research leading to these results has received funding from the European Research
Council under the European Union's Seventh Framework Program (FP7/2007-2013) including ERC grant agreements 240672, 291329, and 306478.
We  acknowledge support from the Brazilian Instituto Nacional de Ci\^encia
e Tecnologia (INCT) e-Universe (CNPq grant 465376/2014-2).

This manuscript has been authored by Fermi Research Alliance, LLC under Contract No. DE-AC02-07CH11359 with the U.S. Department of Energy, Office of Science, Office of High Energy Physics.
\bibliography{trial}

\end{document}